\documentclass{article}

\usepackage[T1]{fontenc}
\usepackage[utf8]{inputenc}
\usepackage{sourceserifpro}
\usepackage{graphicx}
\usepackage[section]{placeins}
\usepackage[export]{adjustbox}
\usepackage[parfill]{parskip}
\usepackage{amsmath}
\usepackage{amsfonts} 
\usepackage{hyperref}  
\usepackage{authblk}
\usepackage{cleveref}
\hypersetup{pdfborder={0 0 0} } 

\usepackage[a4paper, total={6in, 9in}, footskip=0.3in]{geometry} 
\usepackage[
backend=biber,
style=authoryear-comp,
]{biblatex}
\usepackage{todonotes}
\usepackage{xcolor}

\usepackage{setspace}

\usepackage{lastpage}
\usepackage{fancyhdr}

\usepackage{datetime}
\newdateformat{monthyeardate}{%
  \monthname[\THEMONTH], \THEYEAR}

\usepackage{enumitem}

\usetikzlibrary{fit,matrix,positioning}

\title{\textbf{Artificial Utopia: Simulation and Intelligent Agents for a Democratised Future}}

\author[1]{Yannick Oswald\thanks{corresponding author: y-oswald@web.de}}

\affil[1]{Institute of Sustainability and Geography , University of Lausanne}

\date{\monthyeardate\today}
\begin{document}

\maketitle

\begin{abstract}
Prevailing top-down systems in politics and economics struggle to keep pace with the pressing challenges of the 21\textsuperscript{st} century, such as climate change, social inequality and conflict. Bottom-up democratisation and participatory approaches in politics and economics are increasingly seen as promising alternatives to confront and overcome these issues, often with  'utopian' overtones, as proponents believe they may dramatically reshape political, social and ecological futures for the better and in contrast to  contemporary authoritarian tendencies across various countries. Institutional specifics and the associated collective human behavior or culture remains little understood and debated, however. In this article, I propose a novel research agenda focusing on 'utopian' democratisation efforts with formal and computational methods as well as with artificial intelligence -- I call this agenda 'Artificial Utopia'. Artificial Utopias provide safe testing grounds for new political ideas and economic policies 'in-silico' with reduced risk of negative consequences as compared to testing ideas in real-world contexts. An increasing number of advanced simulation and intelligence methods, that aim at representing human cognition and collective decision-making in more realistic ways, could benefit this process. This includes agent-based modelling, reinforcement learning, large language models and more. I clarify what some of these simulation approaches can contribute to the study of Artificial Utopias with the help of two institutional examples; the citizen assembly and the democratic firm. 

\end{abstract}

\section{Introduction}\label{sec:intro}

The evolution of human societies has been profoundly shaped by our capacity for collective decision-making and the development of complex cultural, political and economic systems \parencite{brinkmann2023machine, carballo2014cooperation} and it will continue to shape our future. Historically, at least since the emergence of Western empires, top-down governance structures have dominated \parencite{mcneill1963rise}, but they evidently struggle to address multifaceted collective coordination challenges such as climate change, social inequality, and geopolitical conflicts. In response, there is a growing interest in bottom-up democratization and participatory approaches, which advocate for decentralized decision-making processes encompassing both political institutions as well as economic decision-making. 
Despite the theoretical appeal of participatory systems, our understanding of the specific institutional frameworks and collective behaviors that facilitate or hinder their success remains limited. To address this gap, in this work, I propose a novel research agenda termed Artificial Utopia, which involves computer simulations of alternative economic and democratic political systems, exploring the nature of bottom-up collective decision-making in novel and innovative contexts. Additionally, I outline concrete proposals for building bridges between related social sciences and computational approaches. I argue that alternative and improved democratic systems, systems that go beyond electoral representative democracy, are important objects of scientific inquiry in the context of this agenda. I pay particular attention to not only collective decision-making but also to ensuing challenges, and I discuss how modern simulation and artificial intelligence (AI) methods aid our understanding of those challenges and probe them for a wide range of traits and dynamics. Lastly, I speculate on the future of Artificial Utopia research, related methodological questions and ethical considerations. 

I begin by motivating the research field of Artificial Utopia in more detail. The 21\textsuperscript{st} century presents humankind with unprecedented challenges. Almost every aspect of human life has greatly accelerated. Population and resource consumption have grown exponentially over the past 200 years \parencite{steffen2015trajectory, rockstrom2024planetary}. Several indicators improved globally, like average life expectancy \parencite{dattani2023life}, but immense ecological pressures, such as climate change have emerged, too. Various studies demonstrate that current human civilization is not sustainable and destroys the ecological foundations of life including biodiversity, atmospheric and oceanic health \parencite{richardson2023earth, gupta2024just}. At the same time, inequalities are wide-ranging, both in terms of economic wealth and responsibility for environmental impact \parencite{oswald2020large, chancel2022global}. No country currently achieves decent social and health outcomes while using resources sustainably \parencite{o2018good}. 
This tension between sustainability and human well-being has led to growing scrutiny of the existing economic and political systems that civil society depends on to meet human needs. Beyond social and ecological realms, geopolitical crises and wars are escalating and it is not clear whether the international community will ever cooperate to a degree sufficient to resolving ecological breakdown \parencite{scheffran2023limits}. On the contrary, several previously democratic nations worldwide are turning to authoritarian forms of government, and political parties with fascist ideological elements on their agenda are on the rise \parencite{waldner2018unwelcome}. In response to these intertwined crises - ecological, social, and political - many scholars and practitioners are increasingly turning to bottom-up utopian democratisation efforts as a means to reimagine governance, redistribute power, and create participatory systems that can navigate the challenges of the 21\textsuperscript{st} century \parencite{durand2024planning, steinberger2024democratizing}.

Although it can be challenging to define utopia, it is generally seen as a society in which everything works better as compared to the above described status quo, and does so through a reinvention of how society functions at the governmental and economic level \parencite{giroux2003utopian}. 
Some scholars have explicitly argued that applying utopian thinking to the social and ecological crises of the 21\textsuperscript{st} century is valuable and can help shape our vision for humanity’s future \parencite{zuk2020role}. Of course, utopian re-imagination of society is not new or unique to the 21\textsuperscript{st}  century. Conceptualizations of 'utopia' span a broad historical spectrum, from antiquity and Plato’s early philosophical visions in 'The Republic' \parencite{plato2004republic} to early twentieth-century thinkers like Bertrand Russell and his 'Proposed Roads to Freedom' \parencite{russell1918proposed}. Today, alternative political and economic systems are once again at the forefront of discussions and often espouse an explicitly utopian character \parencite{schmelzer2022future, bastani2019fully, bregman2017utopia}. At the same time, many questions around decision-making in these systems remain and there is no coherent framework for assessing them and little clarity on what tools to use to do so. Can radical and sweeping changes to society really work? Which proposals are feasible? Are there fundamental limits to 'utopian' ideas that arise from decision-making challenges? How can we ensure novel forms of social and economic organisation are reliably democratic and participatory? And how can we study these questions systematically, what scientific and computational tools are suited for what aspect? 

For the purpose of tractability and having a meaningful analytical lens, I focus on two specific kinds of widely proposed, and partially implemented, democratic decision-making systems that are often associated with utopian futures: (i) in politics; the citizen assembly and (ii) in economics; the democratic firm. Both are intuitive entry points to utopia since they constitute quite radical alternatives compared to existing 'mainstream' democratic institutions, but also are practiced by several communities and collectives. The departure from established institutions is in explicitly participatory and deliberative processes that go beyond the interplay of competitive political parties and the voter. People are meant to become active agents of change with a much deeper engagement in decisions about the purpose of political activities and the allocation and distribution of resources \parencite{caluwaerts2016generating}.

\section{From Social Choice to Artificial Utopia - a research agenda}\label{AUintro}

Traditional mathematical models of democratic processes include social choice theory and public choice theory \parencite{iversen2018redistribution, list2013social}. While these theories uncovered various important features of voting systems like  Arrow's impossibility theorems or the Condorcet paradox, they rely on idealised and axiomatic assumptions about political systems \parencite{maskin2014arrow, gehrlein1983condorcet}. Social and public choice theory often employ equilibrium-oriented models \parencite{goodman2004political, myerson2013fundamentals}. Similarly, collective behavior models in economics, such as studies in opinion formation on networks or studies of 'herd' behavior, focus on mathematically tractable dynamics that converge to specific equilibria \parencite{golub2010naive, banerjee1992simple}. But equilibrium models naturally limit the scope of analysis because often they either do not specify explicit dynamic processes, which is critical when understanding how collective decisions unfold \parencite{pangallo2024equations}, or, if they do, assumptions about the behavior of agents are purposefully so restrictive that the dynamic process exhibits an analytically tractable character. This is why they fail to grasp the complex, 'messy' and highly dynamic nature of real-world democratic processes and, therefore, are not well-suited to imagine new forms of political and economic democracy that go substantially beyond the status quo. Because precisely when testing new ideas, it is important to scope a wide range of behavioural outcomes, which can arise from fluctuations and strong heterogeneity of people's behavior and values.

In contrast, computational simulations are increasingly employed to explore explicit alternatives to existing political and economic systems. For instance, some computational work challenges the notion of 'capitalist realism' - the assertion that most political and economic scientific models operate within a fixed capitalist logic \parencite{pahl2023envisioning}. This work intends to test radical changes to politics and economy, which lie outside of a capitalist economic logic, but does so in computational simulations, rather than in real-world experiments. The idea is that then one does so in an environment with substantially less risk of negative consequences \parencite{pahl2023envisioning}. In a related study, \textcite{gerdes2023commonsim} simulate a hypothetical 'utopian' society in which small-scale communities plan their economic production and trade with other communities in a 'needs-based' ex-ante framework, independent of conventional economic markets. Earlier, \textcite{almudi2017economics} studied distinct ideologies competing for different kinds of future utopias. 
However, from these studies, few implications for policy are deduced. Some work approaches novel democratic systems in more specific ways using agent-based simulation and also linking to existing structures. \textcite{carpentras2024empowering} study participatory budgeting designs and find that lessons from 'collective intelligence research' can aid particularly in the inclusion of minority groups when solving policy problems. Another recent work investigates consensus formation in citizen assemblies in a polarized society via multi-agent models \parencite{barrett2024beyond}. The advantage of computational approaches is therefore twofold: First, they allow for the synthesis and testing of hypotheses within artificial settings, enabling simulations of political and economic dynamics to test a multitude of hypotheses that may be impractical or too slow to develop and to evaluate in real-world contexts. And moreover, they prove well-suited for theorizing \textit{why}, that is under what conditions, democratisation efforts succeed or fail.

What is still missing from the computation-oriented discourse is however at least five-fold: (i) there is no integrated theoretical framework guiding research efforts, (ii) there is little clarity on what kind of simulation approaches to employ for what kind of research questions; what are the appropriate domains of validity for different approaches?, (iii) there is little engagement with, or learning from, existing or past egalitarian societies, (iv) the transition dynamics from a status quo to a 'utopia' society are underexplored and (v) there is little clarity on what challenges emerge during unorthodox democratisation efforts and how to overcome them.

I propose that research focuses on closing these knowledge gaps, and I suggest a partial agenda. The principal idea is to follow diverse proposals of democratisation \parencite{hadfi2022augmented, durand2024planning, steinberger2024democratizing} and test their claims. This agenda might draw inspiration from alternative political, economic and social modes that so far are studied predominantly outside of the computational sciences. For example, there is a wealth of ideas on how societies can manage to be economically and politically democratic and egalitarian in anthropological research \parencite{graeber2021dawn, boehm2009hierarchy}, but to date this has inspired few simulation efforts. Most anthropologically inspired simulation-based research focuses on explaining observed behaviour in existing tribal and marginalised societies \parencite{power2017social, bird2015prosocial}, but not so much on transferring ideas for the purpose of reinventing society and testing them at scale. Nonetheless, on top of that, I suggest aiming for generating as much theory-empiry synergy as possible and therefore also advocate for pragmatism when choosing research topics. 

Following this pragmatic line of thinking, two prominent ideas that stand out among democratisation efforts are citizen assemblies and democratic firms, the latter which includes worker cooperatives. This is because they are not only theoretical in nature but have been successfully implemented several times across countries and cultures. The related literature in case studies is relatively rich \parencite{devaney2020ireland, elstub2021scope, burdin2009new, itten2022digital, majee2009building, king2023local}. Furthermore, several activist movements and organisations advocate for them. Citizen assemblies, for example, are a key demand by the climate movement Extinction Rebellion 
\parencite{berglund2020reimagining}. Democratic firms exist world-wide. While still a minority-type of firm, some, like the Spanish Mondragon Corporation, are large multi-national enterprises \parencite{bretos2017challenges}. This ambiguous status, as a partially existing 'minority institution', and as a radical alternative to predominating modes of political and economic decision-making, makes citizen assemblies and democratic firms uniquely suited to be studied in the context of 'Artificial Utopia'. Both allow for empirical insights based on existing real-world case-studies and consequently for comprehension of their character at the individual and collective level, while also still being largely obscure from a computational and theoretical perspective. Moreover, both are sufficiently generic in their flat organisational structure and internal deliberation processes that lessons learned might be generalized to other democratisation efforts, including developments in digital and augmented democracy \parencite{gudino2024large, hadfi2022augmented, pournaras2020proof} or democratic planning \parencite{durand2024planning}. This way, theories of 'utopian' democratisation can be developed alongside empirical foundations. Indeed, while there are some past studies of democratic firms in economics, similarly to social choice theory, they employed idealised equilibrium models largely without grounding their assumptions in empirical knowledge \parencite{bowles1993political} and without representing complex temporal dynamics. Instead, the minimal requirement for an Artificial Utopia - which I define as a useful simulation framework for studying these institutions in a societal context - is to include those features. 

One option then to study 'utopia' via simulation, but far from being limited to, is to focus on the following exemplary set-up: I suggest a citizen assembly representation with different levels of possible decision-making capability. In reality, most currently existing assemblies only consult governance procedures but are not executive organs -- in simulations that could change. Moreover, such a framework can include the distinct stages of a citizen assembly with (a) the selection procedure of citizens, (b) the selection and presentation of political issues and (c) the decision-making process. A similar distinction of stages has been made in \textcite{barrett2024beyond} but it has only been partially simulated. For modelling step (a), it is possible to employ known algorithms to select 'artificial citizens' from a population \parencite{flanigan2021fair}. Crucially, the economy should be understood and represented as a complex evolving system. Therefore I also suggest to apply this understanding to democratic firms specifically. They could be integrated into an evolutionary model of the economy where they must compete with conventional hierarchical firms in order to change the overall economic paradigm. For this purpose, a Schumpeterian and Keynesian agent-based model is possibly ideal \parencite{dosi2019more}. The Schumpeterian school emphasizes evolutionary bottom-up dynamics with firms and entrepreneurs as creative (and destructive) forces \parencite{mellacher2021wage}. The Keynesian school, complementarily, emphasizes government intervention and monetary policy, which can incentivize or even strongly support certain transition processes.

\begin{figure}[hbt!]
\centering
 \includegraphics[width=10cm, height=7cm]{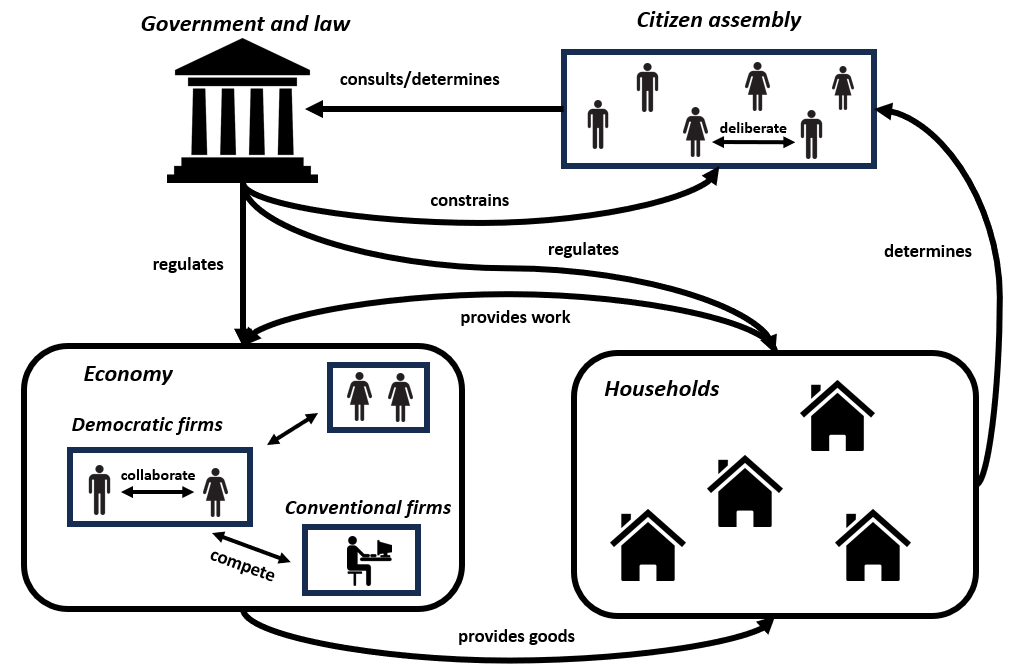}
  \caption{Artificial Utopia concept}
  \label{Figure 1: Artificial Utopia framework and concept}
   \noindent
   \begin{minipage}{0.9\textwidth}
  This figure presents a conceptual high-level presentation of a potential Artificial Utopia. In this specific instance, citizen assemblies are implemented as a governance organ, while democratic firms compete against conventional firms in a market-economy. This figure is only supposed to be indicative and does not necessarily specify model architecture. 
  \end{minipage}
  \medskip
\small
\end{figure}

While specifics should belong to defined research efforts, here I outline representation options. In Figure \ref{Figure 1: Artificial Utopia framework and concept}, for example, I show how both citizen assemblies and democratic firms could come together in a holistic understanding of alternative societies. Going beyond that, in the following section, I will discuss in more detail what kind of simulation approaches could enable specific aspects of Artificial Utopia to become fruitful research ground.

\FloatBarrier

\newpage
\section{Mapping Simulation Approaches to Artificial Utopia}\label{map}

Looking back at the pioneering work on artificial societies by \textcite{epstein1996growing} and the social context of (artificially) intelligent agents by \textcite{conte1995cognitive}, advancements in computational power and methods have transformed the ability to analyze and simulate societies. Digital twins of entire cities are an established part of modern economic geography \parencite{juarez2021digital} and digital twins of political and democratic processes are now in an initial development phase \parencite{garcia2024digital}, to name just a few. Tools such as multi-agent simulations, AI, and big data now allow for unprecedented sophistication in representing human behavior and, potentially, for substantial progress in science overall -- a development that also has been termed `Simulation Intelligence' \parencite{lavin2021simulation}. Of course, there are many persisting limitations, especially when representing human cognition \parencite{wang2025limits}, but nevertheless the range of computational methods certainly has grown substantially. In particular, these advancements provide a promising foundation for studying utopian democratic concepts. In this section, I discuss a selection of computational simulation approaches and how they relate to the Artificial Utopia agenda and to citizen assemblies and democratic firms as examples of bottom-up democratic institutions.

\subsection{Selection of Simulation Approaches}\label{selectionofsimulation}

The following modelling approaches were selected for their relevance and potential to explore complex dynamics inherent in any given Artificial Utopia and dynamics relating to citizen assemblies, democratic firms or related questions: (i) Computational and algorithmic game theory, including stochastic and quantum game theory, (ii) Reinforcement learning and Deep reinforcement learning, (iii) Large Language Models (LLMs), (iv) Agent-Based Modelling and (v) System Dynamics.  

This list is not comprehensive, but it covers most of the popular cutting-edge approaches to improve models of human individual and collective behavior. Any of these methods has particular strengths and weaknesses and characteristic domains of validity. I first describe these approaches and their use cases and then map them in section \ref{mappingtheissues} to challenges arising in citizen assemblies and democratic firms identified in the literature.

\subsection{Computational Game Theory}\label{GT}

Game theory is a mathematical framework for analyzing strategic interactions. Traditional game theory explores competitive scenarios where achieving consensus is against individual rationality. For example, it studies zero-sum games where the pay-offs do not motivate pursuing consensus or collaboration \parencite{nash1950}. This classical perspective could be still essential for revealing precisely which situations in deliberation procedures could be perceived as zero-sum. The focus of game theory on incentives in interactions makes it, despite its focus on equilibrium mathematics, a possible starting point to conceptualise and analyse negotiation and deliberation processes in novel settings.

Stochastic and quantum game theory go beyond these classical perspectives by incorporating complex decision-making features. Stochastic approaches allow for uncertainty in strategies, reflecting real-world human behavior \parencite{camerer2003}. Quantum game theory introduces concepts like the superposition of opinions, potentially representing where individuals hold ambivalent or mixed feelings, and emotional entanglement which could model interpersonal connections that synchronize emotional responses \parencite{eisert1999, trisetyarso2024quantum}. However, the number of applications of quantum game theory so far is low which could reflect a large potential for future research or larger hurdles to find suitable applications.

At the very least, these methods extend the classical paradigm to include more complex behavioral considerations, enabling different perspectives on group dynamics and the evolution of cooperation \parencite{axelrod1984, zhang2022game, joshi2024macroeconomic, myerson1991}. Importantly, from a computational perspective, they may also open up new possibilities for algorithmic game theory to probe complex interactions for specific equilibria or inherent instability, especially when considering repeated games over many time periods, which then also opens up dynamic deliberation processes to game-theoretical analysis \parencite{mertens1990repeated}. 

\subsection{Agent-based modelling}\label{ABM}

While game theory often focuses on few agents or bilateral interactions, agent-based modelling (ABM) is a general purpose bottom-up simulation approach that is able to represent large numbers of people and which is increasingly applied in economics and politics \parencite{axtell2022agent, de2014agent}. Among other things, it has been applied to consensus finding, negotiation and spread of opinions \parencite{luo2008agent}. ABM is very well-suited if interactions and social networks between agents play a role in shaping aggregate emergent outcomes despite agents being 'locally informed' decision-makers. Yet, there are few studies that tackle concrete questions around socio-political and economic-political institutions, particularly when confronting the possibility of reimagining the status quo and engaging in the search for alternatives. Common critiques to agent-based modelling include that it is difficult to verify them empirically and that they are computationally expensive \parencite{an2021challenges}. Improving empirical verification and validation however has become a primary research focus in ABM \parencite{pangallo2024data}. Overall, the potential for ABM to explore utopian economics and politics seems vast because one can build 'tangible' model worlds, and depending on the simulation interface observe them evolve in real-time (for example in Netlogo).

\subsection{Reinforcement learning}\label{RL}

A key limitation of traditional agent-based modelling (ABM) is its reliance on simple, stochastic rules to represent human decision-making, such as fixed savings rates in an economic model for example. While these heuristics mirror theoretical assumptions, they often fail to capture the adaptive, pattern-analyzing nature of human behavior \parencite{an2021challenges}. Reinforcement learning (RL) is an approach suited to overcoming this limitation by making agents learn behaviors through interaction with their environment. Unlike pre-programmed rules, RL specifies goals and objective functions that agents aim to maximize, leading to emergent, goal-driven behaviors. 

Deep reinforcement learning (Deep RL), which combines RL with neural networks, has achieved remarkable success in domains like chess, Go, and video games and training large language models \parencite{mnih2015human, silver2016mastering}. Moreover, RL mirrors mechanisms of human learning in neuroscience \parencite{botvinick2020deep}. Applied to ABM, Deep RL modelled cognitive mechanisms in classic frameworks like Sugarscape and Schelling’s segregation model \parencite{jager2021using}. Deep RL has been further applied to public good problems \parencite{barfuss2019deterministic, strnad2019deep} and taxation models, where government agents optimize policies while workers optimize income strategies \parencite{zheng2022ai}. These innovations may enhance ABM’s ability to simulate complex societal dynamics realistically. Of course, there remain substantial methodological challenges in applying reinforcement learning to complex multi-agent settings such as understanding the emergent dynamics of a non-stationary system with many learning and adaptive agents \parencite{barfuss2022dynamical} which is particularly important for agent alignment and desired aggregate system properties such as sustainability, for example.  

\subsection{Large Language Models (LLMs)}\label{LLMs}

Large Language Models are probably by far the most prominent technology in recent years \parencite{guo2024}. Since the introduction of ChatGPT in 2022, they have sparked both optimism and alarm on their methods and capabilities. In the case of the computational simulation of democratisation processes, they might prove very useful. 

LLMs excel in natural language understanding and generation, enabling them to simulate linguistically sophisticated agents. This capability can also enhance agent-based models (ABMs) by introducing conversational dynamics between agents and perhaps improve specifically political and economic investigations where language plays an important role \parencite{gao2024large}. Indeed, LLMs already can simulate inter-agent political dynamics in settings such as the US senate \parencite{baker2024} or are used in 'augmented democracy' for providing 'expert peers' to citizens that supposedly enable them to make more informed decisions in democratic processes \parencite{gudino2024large}. 

LLMs also are increasingly used to study the emergence of cooperation and sustainable or unsustainable behavior \parencite{piatti2024cooperate}. For example, they have been applied to study climate negotiations \parencite{zhang2022ai}, which could be transferable to citizen assemblies or a democratic firm context. By incorporating LLMs, ABMs may achieve more realistic simulations of democratic processes but come at a high computational expense, specifically during the machine-training process. Furthermore, there are many methodological limitations and risks associated with large language models, but our goal here primarily is to scope what they could contribute to scientific discovery in Artificial Utopias. For a comprehensive assessment of risks emerging from sophisticated agents, like large language models, the reader might refer to \textcite{hammond2025multi}. Directly relevant to this perspective here is, however, the issue of value alignment in LLMs \parencite{khamassi2024strong}. Because LLMs are trained on text data from specific contemporary cultures, it is questionable how well LLM agents can represent change in values and culture. More specifically, it is an unsolved problem how, and if at all, LLMs can represent a value and culture shift that a citizen assembly or cooperative firm might cause. Are they capable of representing the change of values of an individual person as well as social transformation? And along which of their properties would we measure such transformation and how does that map to the analysis of humans?

\subsection{System Dynamics}\label{SD}

System Dynamics (SD) is a numerical equation-based approach. It is appropriate for analysing systems that can be divided into stocks and flows of quantities and feedbacks between those quantities. In our context, SD constitutes a complementary aggregate-level approach to the other more individual-focused approaches. 

Indeed, within political science or economics, SD so far has mostly been applied to tackle questions in macro-scale systems like international relations \parencite{fisunouglu2019system}, global political dynamics \parencite{root2017global} and macro-economics \parencite{radzicki2020system}. SD can also be coupled with micro- and meso-level approaches such as ABM for a holistic system analysis \parencite{martin2015combining}. We explore below how SD could contribute to the study of utopias and citizen assemblies and democratic firms in particular.

\begin{figure}[hbt!]
\centering
 \includegraphics[width=13cm, height=7cm]{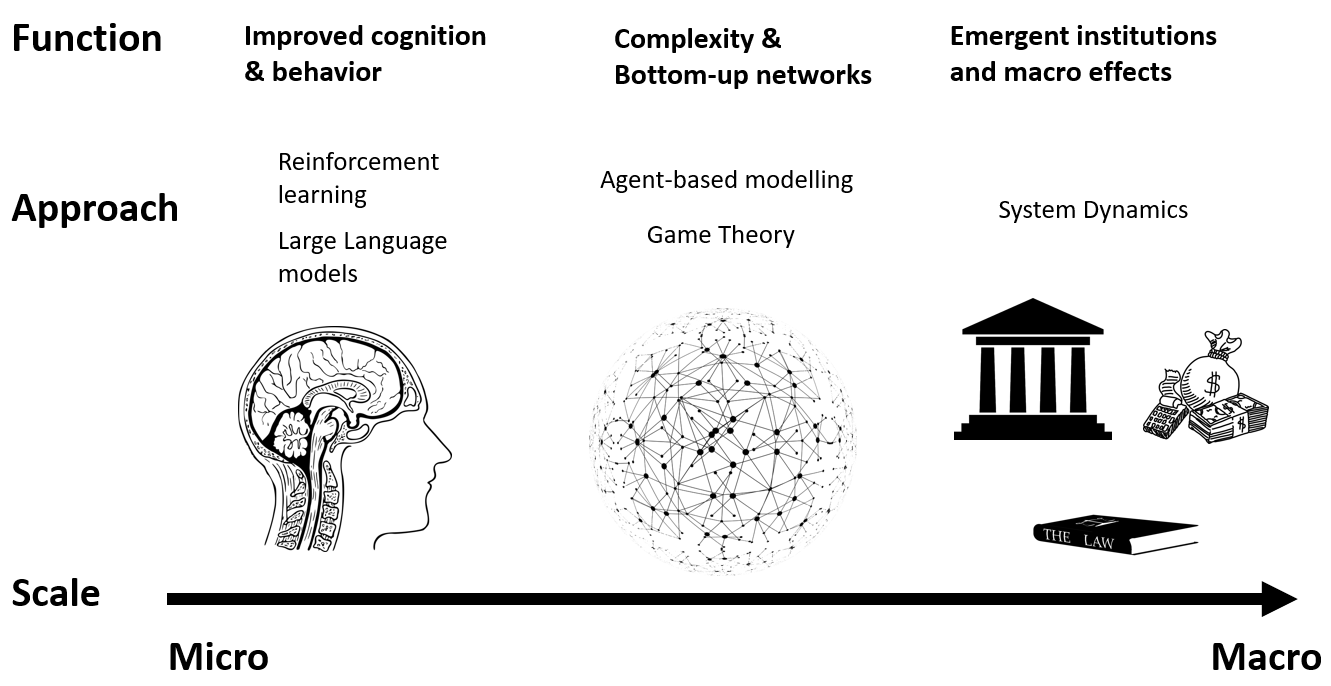}
  \caption{Simulation approaches classification:}
  \label{Figure 2: Simulation approaches classification}
  \medskip
  \noindent
  \begin{minipage}{0.9\textwidth}
  This figure presents a systematization of distinct modelling approaches according to function and scale. Although these features characterise certain domains of validity, none of the domains are absolute. Approaches like Large Language Models are considered micro-level in this context, because we consider them for the application of enhancing models of human cognition in agent-models. In other research domains, large language models may have entirely different purposes. This is similar for the other approaches, too.
  \end{minipage}
\small
\end{figure}

Figure \ref{Figure 2: Simulation approaches classification} summarises the introduced simulation approaches and systematises them according to function, approach and scale. In terms of function, reinforcement learning and LLMS are focused on more realistic representations of human cognition and behavior, ABMs and game theory on interactions and group dynamics, and system dynamics on emergent effects and institutions. 

\FloatBarrier

\subsection{Mapping Potential Issues in Citizen Assemblies and Democratic Firms to Simulation Approaches}\label{mappingtheissues}

In this section, I map the simulation approaches to challenges identified arising in citizen assemblies and democratic firms and which seem suitable to be theorized with the help of computational simulations. In doing so, I outline a set of pathways for advancing the study of Artificial Utopias. I focus on challenges arising in those systems because, despite the fact that citizen assemblies and democratic firms are found to work quite well in many contexts, it is particularly of policy relevance to enhance our understanding of what might go wrong in either of the two cases and within democratisation efforts at large. This way, institutional designs can be made more robust and resilient against anti-democratic forces and systemic shocks. 

\subsubsection{Identified challenges}

Citizen assemblies are complex multi-stage procedures with challenges at each stage. The first stage includes defining what issue a citizen assembly negotiates and what the desired outcome is, as well as the selection of citizens. This stage can be subject to manipulation through influential stakeholders and be subject to selection biases, for example, self-selection bias in participants \parencite{flanigan2021fair, flanigan2024manipulation, stadelmann2016exclusive}. The next stage, in which participants learn and deliberate about the policy issue of interest, might not always be effective or inclusive. For instance, it has been argued that aiming at consensus for a policy decision could be detrimental, given diverse underlying needs of the population and given that some groups might be less prone to defending their position than others \parencite{machin2023democracy}.  Even if consensus is achieved, it can also hint at a sub-optimal or ideological understanding of the problem that is being deliberated on if sufficiently diverse preferences are assumed \parencite{mann2024agents}. In one experimental study, the citizen assembly process even led to increased polarization of political views and also of views on specific issues \parencite{gershtenson2010creating},  which is opposed to the many speculative arguments that citizen assemblies will lead to decreased polarization \parencite{ejsing2023green}. In any case, the effectiveness of deliberation seems to be highly sensitive to group composition and participants' profiles \parencite{suiter2016deliberative, kuntze2021citizen}. Moreover, the role of internal factors such as emotions, personal networks, and power structures in the deliberation process is not well understood and warrants further study \parencite{o2002rhetoric}. Lastly, the role of external events on the deliberation process, such as current political sentiment or major events, is understudied which the lack of studies on this topic demonstrates.

Democratic firms face many similar challenges because their governance form is democratic deliberation. However, they operate distinctly from citizen assemblies because they are competing in market economies which exert pressure to perform, and their key performance indicator is not only an emerging consensus but financial viability. Therefore, as long as the overall economic system logic remains unchanged, one of the biggest challenges for democratic firms is simply to survive in competitive markets \parencite{mellacher2021wage}. That said, existing literature provides qualitative insights into the internal operations and why democratic firms may be outcompeted by regular firms or not. Sometimes, democratic firms actually derive competitive advantages, for example, from knowledge sharing among each other \parencite{basterretxea2011management, basterretxea2012impact}. Among the reasons they fail to remain competitive are information processing failures, which include, for instance, not shutting down unprofitable departments or products, emergent unintended hierarchies \parencite{basterretxea2022corporate},and slow eco-innovation processes \parencite{basterretxea2024eco}. It also has been observed that democratic firms do not achieve optimal team composition because instead of hiring the most qualified from the job market, there is a tendency to resort to personal networks to recruit new employees, a practice that can lead to nepotism \parencite{basterretxea2019can}. In conclusion, bottom-up democratic systems face various pitfalls that warrant further study to understand the conditions that enable or hinder them.

\subsubsection{Simulating challenges in citizen assemblies and democratic firms}

In a next step, I map above introduced computational simulation approaches to the identified challenges and outline how they might improve our understanding of those processes. Table 1 depicts an indicative and non-exhaustive mapping and elaborates on the reasoning of what each simulation approach could contribute.

In summary, game theory approaches are appropriate for clarifying the role of individual incentives within deliberation processes. Stochastic game theory approaches are naturally good at representing the large uncertainty that is inherent to them and at providing a first environment for repeated multi-agent interactions. More speculatively, quantum game theory might be useful to model the 'coupling' of distinct agents as well as the 'superposition' of a spectrum of opinions, which could be a strength when modelling sudden shifts in opinion or sentiment in a group that is deliberating. The strength of reinforcement learning in AI development and control problems lies in its ability to enable a machine agent to learn adaptively from interactions with its environment. Therefore, with respect to collective democratisation settings, reinforcement learning might be useful in defining an individual's objective and how they learn from their interactions with their 'deliberation peers'. Fundamentally, agents might learn this way how to navigate negotiation processes. Although, so far, it is far from clear how common learning algorithms, for instance Q-learning, can be adapted to dynamic and complex policy negotiations and deliberation procedures. In large parts, an emerging research agenda would be required to focus on such methodological questions. Moreover, since the 'learning policy' can be varied across agents, RL also likely is suited to model diversity of preferences and 'personalities' across agents. Ultimately, the decision options at any given time step are then far less constrained than they would be in classical rule-based agent-based models. A long-standing challenge in computational models is to find the right balance between simplicity and complexity \parencite{sun2016simple, edmonds2004kiss, edmonds2019} but it often has been argued that too many fundamental elements of social processes are abstracted away in social and economic models - such as language \parencite{lustick2009abstractions}. Language embodies the qualitative aspects of human interactions. Language is central in emotional and power dynamics. It shapes not only social dynamics, but political outcomes and economic up or downturns (for example by statements from prominent personalities). Hence, LLMs could prove revolutionary in setting up agents that speak to each other and do not abstract away anymore from this central human aspect. Indeed, first studies implement such language-based multi-agent models \parencite{gao2024large} and even so in deliberation processes \parencite{betz2021natural} which seems a plausible step towards more elaborate political and economic settings.
Agent-based models (ABMs) are well-suited for simulating citizen assemblies and democratic firms because they are bottom-up interactive agent models. In principle, any collective decision-making problem can be simulated using ABMs; the primary question is in the level of empirical validity and detail achieved \parencite{taghikhah2021does}. This question of empirical validity constitutes perhaps the greatest methodological challenge in simulating citizen assemblies and democratic firms. Simulation efforts therefore always benefit significantly from collaboration with empirical research groups and should actively pursue synergies between qualitative and quantitative evidence, as well as theory-building.

The application of System Dynamics to citizen assemblies, democratic firms, and Artificial Utopias may not be immediately as intuitive as that of previous methods. Nonetheless, system dynamics could be particularly valuable when integrated with macroeconomic, ecological,, or policy dynamics that govern macro-level variables — such as population size, ecological reservoirs and resources, institutions, fixed capital stocks, or monetary flows. These dynamics, while external to bottom-up deliberation processes, can influence them and thus constitute opportunities for integrated modelling approaches.

\begin{table}[ht]
\centering
\small
\renewcommand{\arraystretch}{1.2}
\resizebox{\textwidth}{!}{%
  \begin{tabular}{p{3.5cm} | p{2.8cm} | p{2.8cm} | p{2.8cm} | p{2.8cm} | p{2.8cm}}
    \hline
    \multicolumn{6}{c}{\normalsize\textbf{Mapping Potential Challenges in Citizen Assemblies and Democratic Firms to Simulation Approaches}} \\ \hline
    \multicolumn{1}{c}{} & \multicolumn{5}{c}{\textbf{Micro} \hspace{0.5cm} \ensuremath{\xrightarrow{\hspace{12cm}}}  \hspace{0.5cm} \textbf{Macro}} \\ \hline
    \textbf{Citizen Assemblies} & \textbf{Game Theory Approaches (GT)} & \textbf{Reinforcement Learning (RL)} & \textbf{Large Language models (LLMs)} & \textbf{Agent-based Modelling (ABM)} & \textbf{System Dynamics (SD)} \\ \hline

Consensus might not be achievable or might not be optimal \parencite{machin2023democracy, mann2024agents} & 
GT can quantify the trade-offs between individual and collective payoffs in consensus-making. & 
RL can offer explicit modelling of distinct utility functions related to certain world-views. & 
LLMs can represent attitudes and personal views embodied in language. & 
ABM can reveal emergent patterns of group behavior, such as polarization or coalition formation. & 
SD can help study whether consensus is sustainable in the face of external disturbances. \\ \hline

Bias and manipulation in citizen selection and in topics of discussion \parencite{stadelmann2016exclusive, flanigan2024manipulation} & 
GT may clarify incentives and payoffs for participation. & 
RL could represent learning trajectory of an agent before entering the deliberation stage. & 
LLMs can represent biases and manipulation expressed in natural language. & 
ABM can explore how conflicting worldviews may introduce bias into decision-making. & 
SD may represent external factors that bias assemblies one way or the other. \\ \hline

Opinion change and questionable effectiveness of deliberation \parencite{machin2023democracy, mann2024agents} & 
GT can elucidate incentives and payoffs for when participants change their minds. & 
RL could show how individual opinions evolve over time through trial and error in negotiations. & 
LLMs can help identify specific language and reasoning dynamics that sway opinions. & 
ABM can study whether certain group composition and network structures influence the spread of opinion change. & 
SD again can clarify systemic and/or external feedbacks that enable or block effective change of opinions. \\ \hline

Emotions and personal relations influence rhetoric, content, and the deliberation process \parencite{o2002rhetoric} & GT can define how emotions and personal relations influence perceived payoffs and costs in negotiations.
& 
RL can represent emotions in the objectives of agents and how they influence agents' learning dynamics. & 
LLMs may express emotions in language and simulate the interplay of language and personal relations. & 
ABMs can model how emotions are driven through interactions. & 
SD can study feedbacks caused by emotion across the population or the perception of overall sentiment. \\ \hline

Power, intent to control deliberation, and deceit to achieve goals \parencite{holdo2019power, blue2016framing} & 
GT may clarify the payoffs and risks agents are willing to take to gain power or when they cede it.&
RL can specify objective functions that represent power-seeking behavior. & LLMs can help study how power is represented in language. & ABM can study how power is embodied in social networks and enforced through them.
& SD can study feedbacks of external factors that enforce or redistribute power.
\\ \hline

\textbf{Democratic Firms} & & & & & \\ \hline

Information processing failure \parencite{basterretxea2022corporate, varman2004contradictions} & 
GT can analyze pairwise and group-wise incentives to reject information. & 
RL can study how behaviour and goals change in light of new information or obstruct information processing. & 
LLMs can represent information in natural language and represent reasoning processes at the agent level. & 
ABM can model information flow between agents and examine how information changes when disseminated. & 
SD can model system-wide information flows. \\ \hline

Emergent, unintended power structure \parencite{basterretxea2022corporate} & 
GT can clarify incentives of different agents to gain or retain power. & 
RL may represent objectives that have to do with power-seeking behavior. & 
LLMs can help study communication and language aspects in the enforcement of power. & 
ABM can study power structures in social networks. & 
SD can model governance interventions that deal with the distribution of power. \\ \hline

Slow (sustainable) innovation processes \parencite{basterretxea2024eco} & 
GT can analyze cooperative vs. competitive incentives in ideation and execution. & 
RL may align different objectives per agent under one firm-level objective. & 
LLMs may explore explicit communication and reasoning processes and their role in innovation. & 
ABM is suited to model collective ideation and intelligence & 
SD, as always, may model system wide feedbacks that inhibit or enable innovation.\\ \hline

Surviving in competitive economies \parencite{mellacher2021wage} & 
GT can assess cooperation or competition when groups decide a firm's strategy. & 
RL can model objectives at the individual level or the firm-level for navigating the market. & 
LLMs can represent interactive planning processes with explicit communication. & 
ABMs can model markets consisting of many firms. & 
SD can model market-level and exogenous feedbacks that impact performance. \\ \hline

Nepotism tendency \parencite{basterretxea2019can} & 
GT may analyze benefits vs. costs of nepotistic practices. & 
RL can conceptualize motivations for preferring personal contacts in employment. & 
LLMs can analyze communication styles within social networks to detect nepotism. & 
ABM can depict social networks of employees to understand influence. & SD can model policies against nepotism and their effectiveness at the systemic level. \\ \hline

    \end{tabular}
} 
\caption{Mapping of Issues in Citizen Assemblies and Democratic Firms to Appropriate Simulation Methods}
\label{Table1}
\end{table}

\newpage

\section{Open questions and discussion}\label{roadforward}

While the capability to create sophisticated simulations that generate insights is increasing at a rapid pace, this not only provides opportunities for Artificial Utopia research but raises many practical and ethical questions. First, it is obvious that this capability is not equally distributed and increasingly in the hands of private companies. Most advances are generated in technology conglomerates or well-funded start-ups, not in public institutions. And what can we expect from societal research that occurs within technology companies? \parencite{al2024project}. Often this research is guided by an engineering perspective and by growth imperatives \parencite{gordon2003capitalism}: Who makes the most intelligent and most sophisticated models and who can turn it into profit? However, the question of 'What are we doing this for?' - remains inadequately addressed apart from business case purposes and a few notable open-source exceptions \parencite{guo2025deepseek}. This is what Artificial Utopia intends to change; Artificial Utopia articulates the desire to guide simulation intelligence by socially benevolent and human-centric visions and engages with social and economic creativity and reimagination.

Second, the perspective presented here, including the selection of simulation approaches, can be critiqued on the grounds of ontological path-dependencies and contexts. This means that our current possibilities for reimagining politics and the economy are constrained by the current political and economic contexts. And importantly they are also constrained by the simulation approaches that are available, because those come with implicit ontological assumptions \parencite{ni2019ontological}. Assume, for the sake of argument, a blank slate world with no, or drastically fewer, pre-existing simulation and modelling approaches but still with substantial computational capability. Further assume, in such a scenario, a taskforce were allocated to reimagining alternative political institutions and economic mechanisms and so forth; it is at least plausible that the newly envisioned political and economic environments would give rise to a distinct, or at least partially different, set of simulation needs compared to the tools we currently rely on, which have largely evolved from specific historical contexts and moments. For example, System Dynamics has its roots in managerial theory and was primarily developed for the cybernetic control of business processes \parencite{radzicki2008origin}. Game theory has often been refined through applications to conflict scenarios. \parencite{dimand1996history, schelling1980strategy}. Simulation methods are hence at least partially influenced by the purposes they are conceived for. Ultimately, 'utopian' applications might give rise to distinct simulation needs and therefore also distinct tools of exploration which are not captured here.

\begin{figure}[hbt!]
\centering
 \includegraphics[width=13.5cm, height=7cm]{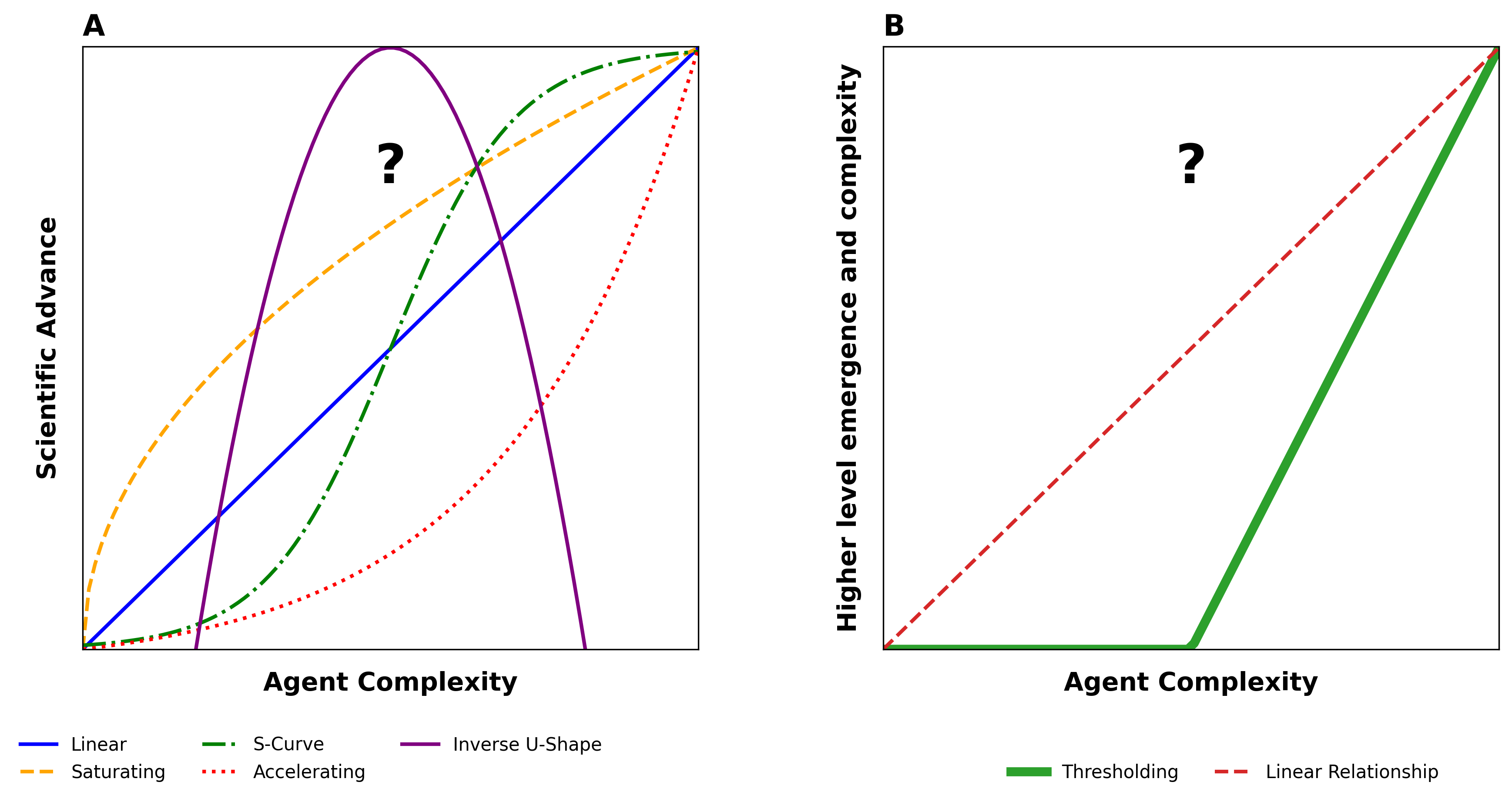}
  \caption{Potential relationships between model sophistication and scientific advance}
    \label{Figure 3: Potential relationships}
  \medskip
    \noindent
    \begin{minipage}{0.9\textwidth}
    This figure presents two central questions in Artificial Utopia research and perhaps social simulation at large. Panel A sketches several potential stylized relationships between the complexity of agents in agent-models and scientific advance on the vertical axis. Neither dimension is precisely specified. Panel B similarly sketches a few possible relationships between the complexity of agent cognition and behavior and higher level emergence. An example for both panels would be the question of whether an agent-based model in economics becomes a better model just because it is a high-dimensional one or whether high-dimensionality at the micro-economic level may even cloud aspects of the aggregate economy? There are no complete answers at this point but I postulate that these are two fundamental questions the agenda of Artificial Utopia should engage with.
    \end{minipage}
\small
\end{figure}

Third, it is not certain a priori whether more sophisticated social simulations serve scientific progress. Indeed, more complex models are of little advantage if the parameter and scenario space simply becomes more uncertain \parencite{puy2022models}. Perhaps the biggest of all lessons from several decades of complexity science is that already simple  behavioral rules can lead to complex and emergent outcomes \parencite{schmickl2016life, schmickl2022strong, mitchell2009complexity}. Examples are the logistic map whose chaotic orbits result in a complex fractal behavior space despite its apparent simplicity \parencite{ausloos2006logistic} and the Game of Life resulting in life-like patterns despite being grounded in extraordinarily simple rules of cell behavior \parencite{conway1970game}. Therefore, with respect to Artificial Utopia, more complex agents enabled by AI do not necessarily lead to improved scientific insight or more complex emergent phenomena. This is also particularly important with respect to sustainability aspects of the research. Training AI applications is energy and resource-intensive and it is not clear whether the 'costs' are worth the insight \parencite{kaack2022aligning}. Figure \ref{Figure 3: Potential relationships} summarizes the uncertainty in the relationship between agent complexity and scientific advance as well as  higher level emergence. 

To conclude, the future possibilities and trajectories for Artificial Utopia research are vast, varied and uncertain. In this article, I have chosen to focus specifically on bottom-up democratization processes, as they form a recurring theme in many scholarly and popular visions of utopia. However, beyond democratisation and universal human rights, this focus should not be misunderstood as limiting the scope of what a utopia could entail. My goal has not been to prescribe a singular vision of utopia but rather to chart a path toward bridging the gap between the aspirations for utopia and our understanding of how such systems might work in practice. By doing so, I hope to provide, or at least refine, tools that allow human societies to collectively steer their future trajectories. 

\section{Funding information}

This project received funding from the REAL-project grant awarded to Prof. Julia Steinberger at the University of Lausanne as well as Prof. Giorgios Kallis and Prof. Jason Hickel at the Autonomous University of Barcelona (UAB) (ERC Grant Number: 101071647). 

\section{Acknowledgments}

I would like to thank Julia Steinberger very much for many helpful discussions related to this work and for feedback on a draft of this work.

\printbibliography

@article{steffen2015trajectory,
  title={The trajectory of the Anthropocene: the great acceleration},
  author={Steffen, Will and Broadgate, Wendy and Deutsch, Lisa and Gaffney, Owen and Ludwig, Cornelia},
  journal={The anthropocene review},
  volume={2},
  number={1},
  pages={81--98},
  year={2015},
  publisher={SAGE Publications Sage UK: London, England}
}

@article{dattani2023life,
  title={Life expectancy},
  author={Dattani, Saloni and Rod{\'e}s-Guirao, Lucas and Ritchie, Hannah and Ortiz-Ospina, Esteban and Roser, Max},
  journal={Our world in data},
  year={2023}
}

@article{basterretxea2022corporate,
  title={Corporate governance as a key aspect in the failure of worker cooperatives},
  author={Basterretxea, Imanol and Cornforth, Chris and Heras-Saizarbitoria, I{\~n}aki},
  journal={Economic and industrial democracy},
  volume={43},
  number={1},
  pages={362--387},
  year={2022},
  publisher={SAGE Publications Sage UK: London, England}
}

@article{mann2024agents,
  title={Agents seeking long-term access to the wisdom of the crowd reduce immediate decision-making accuracy},
  author={Mann, Richard P},
  journal={Philosophical Transactions B},
  volume={379},
  number={1916},
  pages={20220467},
  year={2024},
  publisher={The Royal Society}
}

@article{machin2023democracy,
  title={Democracy, agony, and rupture: A critique of climate citizens’ assemblies},
  author={Machin, Amanda},
  journal={Politische Vierteljahresschrift},
  volume={64},
  number={4},
  pages={845--864},
  year={2023},
  publisher={Springer}
}

@article{flanigan2021fair,
  title={Fair algorithms for selecting citizens’ assemblies},
  author={Flanigan, Bailey and G{\"o}lz, Paul and Gupta, Anupam and Hennig, Brett and Procaccia, Ariel D},
  journal={Nature},
  volume={596},
  number={7873},
  pages={548--552},
  year={2021},
  publisher={Nature Publishing Group UK London}
}

@article{stadelmann2016exclusive,
  title={How exclusive is assembly democracy? Citizens' assembly and ballot participation compared},
  author={Stadelmann-Steffen, Isabelle and Dermont, Clau},
  journal={Swiss Political Science Review},
  volume={22},
  number={1},
  pages={95--122},
  year={2016},
  publisher={Wiley Online Library}
}

@article{o2002rhetoric,
  title={The rhetoric of deliberation: Some problems in Kantian theories of deliberative democracy},
  author={O'neill, John},
  journal={Res Publica},
  volume={8},
  pages={249--268},
  year={2002},
  publisher={Springer}
}

@article{holdo2019power,
  title={Power and Citizen Deliberation: The Contingent Impacts of Interests, Ideology and Status Differences},
  author={Holdo, Markus},
  journal={Journal of Public Deliberation},
  volume={15},
  number={3},
  year={2019},
  publisher={University of Westminster Press}
}

@article{blue2016framing,
  title={Framing and power in public deliberation with climate change: Critical reflections on the role of deliberative practitioners},
  author={Blue, Gwendolyn and Dale, Jacquie},
  journal={Journal of Deliberative Democracy},
  volume={12},
  number={1},
  year={2016},
  publisher={University of Westminster Press}
}

@inproceedings{flanigan2024manipulation,
  title={Manipulation-robust selection of citizens’ assemblies},
  author={Flanigan, Bailey and Liang, Jennifer and Procaccia, Ariel D and Wang, Sven},
  booktitle={Proceedings of the AAAI Conference on Artificial Intelligence},
  volume={38},
  number={9},
  pages={9696--9703},
  year={2024}
}

@article{scheffran2023limits,
  title={Limits to the Anthropocene: geopolitical conflict or cooperative governance?},
  author={Scheffran, J{\"u}rgen},
  journal={Frontiers in Political Science},
  volume={5},
  pages={1190610},
  year={2023},
  publisher={Frontiers Media SA}
}

@article{berglund2020reimagining,
  title={Reimagining democracy},
  author={Berglund, Oscar and Schmidt, Daniel and Berglund, Oscar and Schmidt, Daniel},
  journal={Extinction Rebellion and Climate Change Activism: Breaking the Law to Change the World},
  pages={59--77},
  year={2020},
  publisher={Springer}
}

@article{steinberger2024democratizing,
  title={Democratizing provisioning systems: a prerequisite for living well within limits},
  author={Steinberger, Julia and Guerin, Gauthier and Hofferberth, Elena and Pirgmaier, Elke},
  journal={Sustainability: Science, Practice and Policy},
  volume={20},
  number={1},
  pages={2401186},
  year={2024},
  publisher={Taylor \& Francis}
}

@article{durand2024planning,
  title={Planning beyond growth: The case for economic democracy within ecological limits},
  author={Durand, C{\'e}dric and Hofferberth, Elena and Schmelzer, Matthias},
  journal={Journal of Cleaner Production},
  volume={437},
  pages={140351},
  year={2024},
  publisher={Elsevier}
}

@article{bretos2017challenges,
  title={Challenges and opportunities for the regeneration of multinational worker cooperatives: Lessons from the Mondragon Corporation—a case study of the Fagor Ederlan Group},
  author={Bretos, Ignacio and Errasti, Anjel},
  journal={Organization},
  volume={24},
  number={2},
  pages={154--173},
  year={2017},
  publisher={SAGE Publications Sage UK: London, England}
}

@article{richardson2023earth,
  title={Earth beyond six of nine planetary boundaries},
  author={Richardson, Katherine and Steffen, Will and Lucht, Wolfgang and Bendtsen, J{\o}rgen and Cornell, Sarah E and Donges, Jonathan F and Dr{\"u}ke, Markus and Fetzer, Ingo and Bala, Govindasamy and Von Bloh, Werner and others},
  journal={Science advances},
  volume={9},
  number={37},
  pages={eadh2458},
  year={2023},
  publisher={American Association for the Advancement of Science}
}

@article{oswald2020large,
  title={Large inequality in international and intranational energy footprints between income groups and across consumption categories},
  author={Oswald, Yannick and Owen, Anne and Steinberger, Julia K},
  journal={Nature Energy},
  volume={5},
  number={3},
  pages={231--239},
  year={2020},
  publisher={Nature Publishing Group}
}

@article{o2018good,
  title={A good life for all within planetary boundaries},
  author={O’Neill, Daniel W and Fanning, Andrew L and Lamb, William F and Steinberger, Julia K},
  journal={Nature sustainability},
  volume={1},
  number={2},
  pages={88--95},
  year={2018},
  publisher={Nature Publishing Group}
}

@article{bowles1993political,
  title={A political and economic case for the democratic enterprise},
  author={Bowles, Samuel and Gintis, Herbert},
  journal={Economics \& Philosophy},
  volume={9},
  number={1},
  pages={75--100},
  year={1993},
  publisher={Cambridge University Press}
}

@article{pahl2023envisioning,
  title={Envisioning post-capitalist utopias via simulation: Theory, critique and models},
  author={Pahl, Hanno and Scholz-W{\"a}ckerle, Manuel and Schr{\"o}ter, Jens},
  journal={Review of Evolutionary Political Economy},
  volume={4},
  number={3},
  pages={445--465},
  year={2023},
  publisher={Springer}
}

@article{gerdes2023commonsim,
  title={COMMONSIM: Simulating the utopia of COMMONISM},
  author={Gerdes, Lena and Aigner, Ernest and Meretz, Stefan and Pahl, Hanno and Schlemm, Annette and Scholz-W{\"a}ckerle, Manuel and Schr{\"o}ter, Jens and Sutterl{\"u}tti, Simon},
  journal={Review of Evolutionary Political Economy},
  volume={4},
  number={3},
  pages={559--595},
  year={2023},
  publisher={Springer}
}

@article{nash1950,
  author = {Nash, John F.},
  title = {Equilibrium points in n-person games},
  journal = {Proceedings of the National Academy of Sciences},
  volume = {36},
  number = {1},
  pages = {48--49},
  year = {1950},
  doi = {10.1073/pnas.36.1.48}
}

@book{axelrod1984,
  author = {Axelrod, Robert},
  title = {The Evolution of Cooperation},
  publisher = {Basic Books},
  year = {1984},
  isbn = {978-0465021215}
}

@book{myerson1991,
  author = {Myerson, Roger B.},
  title = {Game Theory: Analysis of Conflict},
  publisher = {Harvard University Press},
  year = {1991},
  isbn = {978-0674341166}
}

@article{eisert1999,
  author = {Eisert, Jens and Wilkens, Martin and Lewenstein, Maciej},
  title = {Quantum games and quantum strategies},
  journal = {Physical Review Letters},
  volume = {83},
  number = {15},
  pages = {3077--3080},
  year = {1999},
  doi = {10.1103/PhysRevLett.83.3077}
}

@book{camerer2003,
  author = {Camerer, Colin F.},
  title = {Behavioral Game Theory: Experiments in Strategic Interaction},
  publisher = {Princeton University Press},
  year = {2003},
  isbn = {978-0691090399}
}

@article{lavin2021simulation,
  title={Simulation intelligence: Towards a new generation of scientific methods},
  author={Lavin, Alexander and Krakauer, David and Zenil, Hector and Gottschlich, Justin and Mattson, Tim and Brehmer, Johann and Anandkumar, Anima and Choudry, Sanjay and Rocki, Kamil and Baydin, At{\i}l{\i}m G{\"u}ne{\c{s}} and others},
  journal={arXiv preprint arXiv:2112.03235},
  year={2021}
}

@article{gao2024large,
  title={Large language models empowered agent-based modeling and simulation: A survey and perspectives},
  author={Gao, Chen and Lan, Xiaochong and Li, Nian and Yuan, Yuan and Ding, Jingtao and Zhou, Zhilun and Xu, Fengli and Li, Yong},
  journal={Humanities and Social Sciences Communications},
  volume={11},
  number={1},
  pages={1--24},
  year={2024},
  publisher={Palgrave}
}

@article{guo2024,
  author = {Guo, T. and Chen, X. and Wang, Y. and Chang, R. and Pei, S. and Chawla, N. V. and Wiest, O. and Zhang, X.},
  title = {Large Language Model based Multi-Agents: A Survey of Progress and Challenges},
  journal = {arXiv preprint arXiv:2402.01680},
  year = {2024},
  url = {https://arxiv.org/abs/2402.01680}
}

@article{baker2024,
  author = {Baker, Z. R. and Azher, Z. L.},
  title = {Simulating The U.S. Senate: An LLM-Driven Agent Approach to Modeling Legislative Behavior and Bipartisanship},
  journal = {arXiv preprint arXiv:2406.18702},
  year = {2024},
  url = {https://arxiv.org/abs/2406.18702}
}

@article{zhang2022ai,
  title={AI for global climate cooperation: modeling global climate negotiations, agreements, and long-term cooperation in RICE-N},
  author={Zhang, Tianyu and Williams, Andrew and Phade, Soham and Srinivasa, Sunil and Zhang, Yang and Gupta, Prateek and Bengio, Yoshua and Zheng, Stephan},
  journal={arXiv preprint arXiv:2208.07004},
  year={2022}
}

@article{piatti2024cooperate,
  title={Cooperate or Collapse: Emergence of Sustainability Behaviors in a Society of LLM Agents},
  author={Piatti, Giorgio and Jin, Zhijing and Kleiman-Weiner, Max and Sch{\"o}lkopf, Bernhard and Sachan, Mrinmaya and Mihalcea, Rada},
  journal={arXiv preprint arXiv:2404.16698},
  year={2024}
}

@article{axtell2022agent,
  title={Agent-based modeling in economics and finance: Past, present, and future},
  author={Axtell, Robert L and Farmer, J Doyne},
  journal={Journal of Economic Literature},
  pages={1--101},
  year={2022},
  publisher={American Economic Association}
}

@article{de2014agent,
  title={Agent-based models},
  author={De Marchi, Scott and Page, Scott E},
  journal={Annual Review of political science},
  volume={17},
  number={1},
  pages={1--20},
  year={2014},
  publisher={Annual Reviews}
}

@article{an2021challenges,
  title={Challenges, tasks, and opportunities in modeling agent-based complex systems},
  author={An, Li and Grimm, Volker and Sullivan, Abigail and Turner Ii, BL and Malleson, Nicolas and Heppenstall, Alison and Vincenot, Christian and Robinson, Derek and Ye, Xinyue and Liu, Jianguo and others},
  journal={Ecological Modelling},
  volume={457},
  pages={109685},
  year={2021},
  publisher={Elsevier}
}

@article{pangallo2024data,
  title={Data-driven economic agent-based models},
  author={Pangallo, Marco and del Rio-Chanona, R Maria},
  journal={arXiv preprint arXiv:2412.16591},
  year={2024}
}

@article{luo2008agent,
  title={Agent-based human behavior modeling for crowd simulation},
  author={Luo, Linbo and Zhou, Suiping and Cai, Wentong and Low, Malcolm Yoke Hean and Tian, Feng and Wang, Yongwei and Xiao, Xian and Chen, Dan},
  journal={Computer Animation and Virtual Worlds},
  volume={19},
  number={3-4},
  pages={271--281},
  year={2008},
  publisher={Wiley Online Library}
}

@article{fisunouglu2019system,
  title={System dynamics modeling in international relations},
  author={Fisuno{\u{g}}lu, Ali},
  journal={All Azimuth: A Journal of Foreign Policy and Peace},
  volume={8},
  number={2},
  pages={231--253},
  year={2019},
  publisher={Center for Foreign Policy and Peace Research, {\.I}hsan Do{\u{g}}ramac{\i} Peace Foundation}
}

@article{root2017global,
  title={Global political dynamics and the science of complex systems},
  author={Root, Hilton L},
  journal={Non-Equilibrium Social Science and Policy: Introduction and Essays on New and Changing Paradigms in Socio-Economic Thinking},
  pages={97--109},
  year={2017},
  publisher={Springer International Publishing}
}

@article{radzicki2020system,
  title={System dynamics and its contribution to economics and economic modeling},
  author={Radzicki, Michael J},
  journal={System dynamics: Theory and applications},
  pages={401--415},
  year={2020},
  publisher={Springer}
}

@article{martin2015combining,
  title={Combining system dynamics and agent-based modeling to analyze social-ecological interactions—an example from modeling restoration of a shallow lake},
  author={Martin, Romina and Schl{\"u}ter, Maja},
  journal={Frontiers in Environmental Science},
  volume={3},
  pages={66},
  year={2015},
  publisher={Frontiers Media SA}
}

@article{basterretxea2024eco,
  title={ECO-INNOVATION IN WORKER COOPERATIVES AND INVESTOR-OWNED INDUSTRIAL FIRMS: A COMPARATIVE ANALYSIS.},
  author={Basterretxea, Imanol and Fern{\'a}ndez-Sainz, Ana and Guti{\'e}rrez-Goiria, Jorge and Santos-Larrazabal, Josu},
  journal={Revista de econom{\'\i}a mundial},
  number={67},
  year={2024}
}

@article{basterretxea2019can,
  title={Can employee ownership and human resource management policies clash in worker cooperatives? Lessons from a defunct cooperative},
  author={Basterretxea, Imanol and Heras-Saizarbitoria, I{\~n}aki and Lertxundi, Aitziber},
  journal={Human Resource Management},
  volume={58},
  number={6},
  pages={585--601},
  year={2019},
  publisher={Wiley Online Library}
}

@article{mellacher2021wage,
  title={Wage inequality, labor market polarization and skill-biased technological change: an evolutionary (agent-based) approach},
  author={Mellacher, Patrick and Scheuer, Timon},
  journal={Computational Economics},
  volume={58},
  number={2},
  pages={233--278},
  year={2021},
  publisher={Springer}
}

@article{betz2021natural,
  title={Natural-language multi-agent simulations of argumentative opinion dynamics},
  author={Betz, Gregor},
  journal={arXiv preprint arXiv:2104.06737},
  year={2021}
}

@article{lustick2009abstractions,
  title={Abstractions, ensembles, and virtualizations: simplicity and complexity in agent-based modeling},
  author={Lustick, Ian S and Miodownik, Dan},
  journal={Comparative Politics},
  volume={41},
  number={2},
  pages={223--244},
  year={2009},
  publisher={City University of New York}
}

@article{puy2022models,
  title={Models with higher effective dimensions tend to produce more uncertain estimates},
  author={Puy, Arnald and Beneventano, Pierfrancesco and Levin, Simon A and Lo Piano, Samuele and Portaluri, Tommaso and Saltelli, Andrea},
  journal={Science Advances},
  volume={8},
  number={42},
  pages={eabn9450},
  year={2022},
  publisher={American Association for the Advancement of Science}
}

@article{almudi2017economics,
  title={The economics of utopia: a co-evolutionary model of ideas, citizenship and socio-political change},
  author={Almudi, Isabel and Fatas-Villafranca, Francisco and Izquierdo, Luis R and Potts, Jason},
  journal={Journal of Evolutionary Economics},
  volume={27},
  pages={629--662},
  year={2017},
  publisher={Springer}
}

@article{carpentras2024empowering,
  title={Empowering minorities and everyone in participatory budgeting: an agent-based modelling perspective},
  author={Carpentras, Dino and H{\"a}nggli Fricker, Regula and Helbing, Dirk},
  journal={Philosophical Transactions A},
  volume={382},
  number={2285},
  pages={20240090},
  year={2024},
  publisher={The Royal Society}
}

@article{barrett2024beyond,
  title={Beyond the echo chamber: modelling open-mindedness in citizens’ assemblies},
  author={Barrett, Jake and Gal, Kobi and Michael, Loizos and Vilenchik, Dan},
  journal={Autonomous Agents and Multi-Agent Systems},
  volume={38},
  number={2},
  pages={30},
  year={2024},
  publisher={Springer}
}

@article{dosi2019more,
  title={More is different... and complex! the case for agent-based macroeconomics},
  author={Dosi, Giovanni and Roventini, Andrea},
  journal={Journal of Evolutionary Economics},
  volume={29},
  pages={1--37},
  year={2019},
  publisher={Springer}
}

@book{epstein1996growing,
  title={Growing artificial societies: social science from the bottom up},
  author={Epstein, Joshua M and Axtell, Robert},
  year={1996},
  publisher={Brookings Institution Press}
}

@article{juarez2021digital,
  title={Digital twins: Review and challenges},
  author={Juarez, Maria G and Botti, Vicente J and Giret, Adriana S},
  journal={Journal of Computing and Information Science in Engineering},
  volume={21},
  number={3},
  pages={030802},
  year={2021},
  publisher={American Society of Mechanical Engineers}
}

@article{basterretxea2011management,
  title={Management training as a source of perceived competitive advantage: The Mondragon Cooperative Group case},
  author={Basterretxea, Imanol and Albizu, Eneka},
  journal={Economic and Industrial Democracy},
  volume={32},
  number={2},
  pages={199--222},
  year={2011},
  publisher={SAGE Publications Sage UK: London, England}
}

@article{basterretxea2012impact,
  title={Impact of management and innovation capabilities on performance: Are cooperatives different?},
  author={Basterretxea, Imanol and Mart{\'\i}nez, Ricardo},
  journal={Annals of Public and Cooperative Economics},
  volume={83},
  number={3},
  pages={357--381},
  year={2012},
  publisher={Wiley Online Library}
}

@article{al2024project,
  title={Project Sid: Many-agent simulations toward AI civilization},
  author={AL, Altera and Ahn, Andrew and Becker, Nic and Carroll, Stephanie and Christie, Nico and Cortes, Manuel and Demirci, Arda and Du, Melissa and Li, Frankie and Luo, Shuying and others},
  journal={arXiv preprint arXiv:2411.00114},
  year={2024}
}

@article{giroux2003utopian,
  title={Utopian thinking under the sign of neoliberalism: Towards a critical pedagogy of educated hope},
  author={Giroux, Henry},
  journal={Democracy \& Nature},
  volume={9},
  number={1},
  pages={91--105},
  year={2003},
  publisher={Taylor \& Francis}
}

@article{zuk2020role,
  title={On the role of utopia in social thought and social sciences},
  author={{\.Z}uk, Piotr},
  journal={History of European Ideas},
  volume={46},
  number={8},
  pages={1047--1058},
  year={2020},
  publisher={Taylor \& Francis}
}

@article{rockstrom2024planetary,
  title={Planetary Boundaries guide humanity’s future on Earth},
  author={Rockstr{\"o}m, Johan and Donges, Jonathan F and Fetzer, Ingo and Martin, Maria A and Wang-Erlandsson, Lan and Richardson, Katherine},
  journal={Nature Reviews Earth \& Environment},
  volume={5},
  number={11},
  pages={773--788},
  year={2024},
  publisher={Nature Publishing Group}
}

@article{gupta2024just,
  title={A just world on a safe planet: a Lancet Planetary Health--Earth Commission report on Earth-system boundaries, translations, and transformations},
  author={Gupta, Joyeeta and Bai, Xuemei and Liverman, Diana M and Rockstr{\"o}m, Johan and Qin, Dahe and Stewart-Koster, Ben and Rocha, Juan C and Jacobson, Lisa and Abrams, Jesse F and Andersen, Lauren S and others},
  journal={The Lancet Planetary Health},
  volume={8},
  number={10},
  pages={e813--e873},
  year={2024},
  publisher={Elsevier}
}

@article{chancel2022global,
  title={Global carbon inequality over 1990--2019},
  author={Chancel, Lucas},
  journal={Nature Sustainability},
  volume={5},
  number={11},
  pages={931--938},
  year={2022},
  publisher={Nature Publishing Group}
}

@book{bastani2019fully,
  title={Fully automated luxury communism},
  author={Bastani, Aaron},
  year={2019},
  publisher={Verso Books}
}

@book{schmelzer2022future,
  title={The future is degrowth: A guide to a world beyond capitalism},
  author={Schmelzer, Matthias and Vetter, Andrea and Vansintjan, Aaron},
  year={2022},
  publisher={Verso Books}
}

@book{bregman2017utopia,
  title={Utopia for realists: And how we can get there},
  author={Bregman, Rutger},
  year={2017},
  publisher={Bloomsbury Publishing}
}

@article{devaney2020ireland,
  title={Ireland’s citizens’ assembly on climate change: Lessons for deliberative public engagement and communication},
  author={Devaney, Laura and Torney, Diarmuid and Brereton, Pat and Coleman, Martha},
  journal={Environmental Communication},
  volume={14},
  number={2},
  pages={141--146},
  year={2020},
  publisher={Taylor \& Francis}
}

@article{elstub2021scope,
  title={The scope of climate assemblies: lessons from the climate assembly UK},
  author={Elstub, Stephen and Carrick, Jayne and Farrell, David M and Mockler, Patricia},
  journal={Sustainability},
  volume={13},
  number={20},
  pages={11272},
  year={2021},
  publisher={MDPI}
}

@article{burdin2009new,
  title={New evidence on wages and employment in worker cooperatives compared with capitalist firms},
  author={Burdin, Gabriel and Dean, Andr{\'e}s},
  journal={Journal of comparative economics},
  volume={37},
  number={4},
  pages={517--533},
  year={2009},
  publisher={Elsevier}
}

@article{varman2004contradictions,
  title={Contradictions of democracy in a workers’ cooperative},
  author={Varman, Rahul and Chakrabarti, Manali},
  journal={Organization studies},
  volume={25},
  number={2},
  pages={183--208},
  year={2004},
  publisher={Sage Publications}
}

@article{itten2022digital,
  title={When digital mass participation meets citizen deliberation: combining mini-and maxi-publics in climate policy-making},
  author={Itten, Anatol and Mouter, Niek},
  journal={Sustainability},
  volume={14},
  number={8},
  pages={4656},
  year={2022},
  publisher={MDPI}
}

@article{majee2009building,
  title={Building community trust through cooperatives: A case study of a worker-owned homecare cooperative},
  author={Majee, Wilson and Hoyt, Ann},
  journal={Journal of Community Practice},
  volume={17},
  number={4},
  pages={444--463},
  year={2009},
  publisher={Taylor \& Francis}
}

@article{gudino2024large,
  title={Large Language Models (LLMs) as Agents for Augmented Democracy},
  author={Gudi{\~n}o-Rosero, Jairo and Grandi, Umberto and Hidalgo, C{\'e}sar A},
  journal={arXiv preprint arXiv:2405.03452},
  year={2024}
}

@article{pournaras2020proof,
  title={Proof of witness presence: Blockchain consensus for augmented democracy in smart cities},
  author={Pournaras, Evangelos},
  journal={Journal of Parallel and Distributed Computing},
  volume={145},
  pages={160--175},
  year={2020},
  publisher={Elsevier}
}

@inproceedings{hadfi2022augmented,
  title={Augmented Democratic Deliberation: Can Conversational Agents Boost Deliberation in Social Media?},
  author={Hadfi, Rafik and Ito, Takayuki},
  booktitle={proceedings of the 21st international conference on autonomous agents and multiagent systems},
  pages={1794--1798},
  year={2022}
}

@incollection{garcia2024digital,
  title={Digital Twins: On Algorithm-Based Political Participation},
  author={Garc{\'\i}a-Marz{\'a}, Domingo and Calvo, Patrici},
  booktitle={Algorithmic Democracy: A Critical Perspective Based on Deliberative Democracy},
  pages={61--79},
  year={2024},
  publisher={Springer}
}

@article{zhang2022game,
  title={Game theory and the evolution of cooperation},
  author={Zhang, Bo-Yu and Pei, Shan},
  journal={Journal of the Operations Research Society of China},
  volume={10},
  number={2},
  pages={379--399},
  year={2022},
  publisher={Springer}
}

@article{trisetyarso2024quantum,
  title={Quantum simulation of coopetition},
  author={Trisetyarso, Agung and Hastiadi, Fithra Faisal},
  journal={Expert Systems with Applications},
  pages={124461},
  year={2024},
  publisher={Elsevier}

}

@article{joshi2024macroeconomic,
  title={Macroeconomic Policies in a Multiparty Democracy: A Game-Theoretic Approach},
  author={Joshi, Yagha},
  journal={Available at SSRN 4894171},
  year={2024}
}

@article{caluwaerts2016generating,
  title={Generating democratic legitimacy through deliberative innovations: The role of embeddedness and disruptiveness},
  author={Caluwaerts, Didier and Reuchamps, Min},
  journal={Representation},
  volume={52},
  number={1},
  pages={13--27},
  year={2016},
  publisher={Taylor \& Francis}
}

@article{taghikhah2021does,
  title={Where does theory have it right? A comparison of theory-driven and empirical agent based models},
  author={Taghikhah, Firouzeh and Filatova, Tatiana and Voinov, Alexey},
  journal={Journal of Artificial Societies and Social Simulation},
  year={2021},
  publisher={University of Surrey}
}

@article{gordon2003capitalism,
  title={Capitalism's growth imperative},
  author={Gordon, Myron J and Rosenthal, Jeffrey S},
  journal={Cambridge Journal of Economics},
  volume={27},
  number={1},
  pages={25--48},
  year={2003},
  publisher={Oxford University Press}
}

@article{schmickl2016life,
  title={How a life-like system emerges from a simplistic particle motion law},
  author={Schmickl, Thomas and Stefanec, Martin and Crailsheim, Karl},
  journal={Scientific reports},
  volume={6},
  number={1},
  pages={37969},
  year={2016},
  publisher={Nature Publishing Group UK London}
}

@article{schmickl2022strong,
  title={Strong emergence arising from weak emergence},
  author={Schmickl, Thomas},
  journal={Complexity},
  volume={2022},
  number={1},
  pages={9956885},
  year={2022},
  publisher={Wiley Online Library}
}

@article{conway1970game,
  title={The game of life},
  author={Conway, John and others},
  journal={Scientific American},
  volume={223},
  number={4},
  pages={4},
  year={1970}
}

@book{mitchell2009complexity,
  title={Complexity: A guided tour},
  author={Mitchell, Melanie},
  year={2009},
  publisher={Oxford University Press}
}

@book{ausloos2006logistic,
  title={The logistic map and the route to chaos: From the beginnings to modern applications},
  author={Ausloos, Marcel and Dirickx, Michel},
  year={2006},
  publisher={Springer Science \& Business Media}
}

@article{kaack2022aligning,
  title={Aligning artificial intelligence with climate change mitigation},
  author={Kaack, Lynn H and Donti, Priya L and Strubell, Emma and Kamiya, George and Creutzig, Felix and Rolnick, David},
  journal={Nature Climate Change},
  volume={12},
  number={6},
  pages={518--527},
  year={2022},
  publisher={Nature Publishing Group UK London}
}

@article{iversen2018redistribution,
  title={Redistribution without a median voter: Models of multidimensional politics},
  author={Iversen, Torben and Goplerud, Max},
  journal={Annual Review of Political Science},
  volume={21},
  number={1},
  pages={295--317},
  year={2018},
  publisher={Annual Reviews}
}

@article{list2013social,
  title={Social choice theory},
  author={List, Christian},
  year={2013},
  journal={Stanford Encyclopedia of Philosophy}
}

@article{gehrlein1983condorcet,
  title={Condorcet's paradox},
  author={Gehrlein, William V},
  journal={Theory and decision},
  volume={15},
  number={2},
  pages={161--197},
  year={1983},
  publisher={Springer}
}

@book{maskin2014arrow,
  title={The Arrow impossibility theorem},
  author={Maskin, Eric and Sen, Amartya},
  year={2014},
  publisher={Columbia University Press}
}

@article{guo2025deepseek,
  title={DeepSeek-R1: Incentivizing Reasoning Capability in LLMs via Reinforcement Learning},
  author={Guo, Daya and Yang, Dejian and Zhang, Haowei and Song, Junxiao and Zhang, Ruoyu and Xu, Runxin and Zhu, Qihao and Ma, Shirong and Wang, Peiyi and Bi, Xiao and others},
  journal={arXiv preprint arXiv:2501.12948},
  year={2025}
}

@article{sun2016simple,
  title={Simple or complicated agent-based models? A complicated issue},
  author={Sun, Zhanli and Lorscheid, Iris and Millington, James D and Lauf, Steffen and Magliocca, Nicholas R and Groeneveld, J{\"u}rgen and Balbi, Stefano and Nolzen, Henning and M{\"u}ller, Birgit and Schulze, Jule and others},
  journal={Environmental Modelling \& Software},
  volume={86},
  pages={56--67},
  year={2016},
  publisher={Elsevier}
}

@inproceedings{edmonds2004kiss,
  title={From KISS to KIDS--an ‘anti-simplistic’modelling approach},
  author={Edmonds, Bruce and Moss, Scott},
  booktitle={International workshop on multi-agent systems and agent-based simulation},
  pages={130--144},
  year={2004},
  organization={Springer}
}

@article{edmonds2019,
   title = {Different Modelling Purposes},
   author = {Edmonds, Bruce and Le Page, Christophe and Bithell, Mike and Chattoe-Brown, Edmund and Grimm, Volker and Meyer, Ruth and Monta\~{n}ola-Sales, Cristina and Ormerod, Paul and Root, Hilton and Squazzoni, Flaminio},
   journal = {Journal of Artificial Societies and Social Simulation},
   ISSN = {1460-7425},
   volume = {22},
   number = {3},
   pages = {6},
   year = {2019},
   URL = {http://jasss.soc.surrey.ac.uk/22/3/6.html},
   DOI = {10.18564/jasss.3993},
}

@article{goodman2004political,
  title={Political equilibrium and the provision of public goods},
  author={Goodman, John C and Porter, Philip K},
  journal={Public Choice},
  volume={120},
  number={3},
  pages={247--266},
  year={2004},
  publisher={Springer}
}

@article{myerson2013fundamentals,
  title={Fundamentals of social choice theory},
  author={Myerson, Roger B and others},
  journal={Quarterly Journal of Political Science},
  volume={8},
  number={3},
  pages={305--337},
  year={2013},
  publisher={Now Publishers, Inc.}
}

@article{pangallo2024equations,
  title={Equations vs. maps: complexity, equilibrium, disequilibrium},
  author={Pangallo, Marco},
  journal={Available at SSRN 4971148},
  year={2024}
}

@book{graeber2021dawn,
  title={The dawn of everything: A new history of humanity},
  author={Graeber, David and Wengrow, David},
  year={2021},
  publisher={Penguin UK}
}

@book{boehm2009hierarchy,
  title={Hierarchy in the forest: The evolution of egalitarian behavior},
  author={Boehm, Christopher and Boehm, Christopher},
  year={2009},
  publisher={Harvard University Press}
}

@article{power2017social,
  title={Social support networks and religiosity in rural South India},
  author={Power, Eleanor A},
  journal={Nature Human Behaviour},
  volume={1},
  number={3},
  pages={0057},
  year={2017},
  publisher={Nature Publishing Group UK London}
}

@article{bird2015prosocial,
  title={Prosocial signaling and cooperation among Martu hunters},
  author={Bird, Rebecca Bliege and Power, Eleanor A},
  journal={Evolution and Human Behavior},
  volume={36},
  number={5},
  pages={389--397},
  year={2015},
  publisher={Elsevier}
}

@book{dimand1996history,
  title={The history of game theory, volume 1: from the beginnings to 1945},
  author={Dimand, Mary-Ann and Dimand, Robert W},
  year={1996},
  publisher={Routledge}
}

@article{ni2019ontological,
  title={Ontological Politics in a World of Political Ontologies: More Realistic (Human) Agents for the Anthropocene?},
  author={n{\'\i} Aodha, Lia},
  journal={Social Simulation for a Digital Society: Applications and Innovations in Computational Social Science},
  pages={7--20},
  year={2019},
  publisher={Springer}
}

@article{radzicki2008origin,
  title={Origin of system dynamics: Jay W. Forrester and the history of system dynamics},
  author={Radzicki, Michael J and Taylor, Robert A},
  journal={US Department of Energy’s introduction to system dynamics},
  year={2008},
  publisher={System Dynamics Society}
}

@book{schelling1980strategy,
  title={The Strategy of Conflict: with a new Preface by the Author},
  author={Schelling, Thomas C},
  year={1980},
  publisher={Harvard university press}
}

@incollection{mertens1990repeated,
  title={Repeated games},
  author={Mertens, Jean-Fran{\c{c}}ois},
  booktitle={Game theory and applications},
  pages={77--130},
  year={1990},
  publisher={Elsevier}
}

@article{mnih2015human,
  title={Human-level control through deep reinforcement learning},
  author={Mnih, Volodymyr and Kavukcuoglu, Koray and Silver, David and Rusu, Andrei A and Veness, Joel and Bellemare, Marc G and Graves, Alex and Riedmiller, Martin and Fidjeland, Andreas K and Ostrovski, Georg and others},
  journal={nature},
  volume={518},
  number={7540},
  pages={529--533},
  year={2015},
  publisher={Nature Publishing Group UK London}
}

@article{silver2016mastering,
  title={Mastering the game of Go with deep neural networks and tree search},
  author={Silver, David and Huang, Aja and Maddison, Chris J and Guez, Arthur and Sifre, Laurent and Van Den Driessche, George and Schrittwieser, Julian and Antonoglou, Ioannis and Panneershelvam, Veda and Lanctot, Marc and others},
  journal={nature},
  volume={529},
  number={7587},
  pages={484--489},
  year={2016},
  publisher={Nature Publishing Group}
}

@article{waldner2018unwelcome,
  title={Unwelcome change: Coming to terms with democratic backsliding},
  author={Waldner, David and Lust, Ellen},
  journal={Annual Review of Political Science},
  volume={21},
  number={1},
  pages={93--113},
  year={2018},
  publisher={Annual Reviews}
}

@article{botvinick2020deep,
  title={Deep reinforcement learning and its neuroscientific implications},
  author={Botvinick, Matthew and Wang, Jane X and Dabney, Will and Miller, Kevin J and Kurth-Nelson, Zeb},
  journal={Neuron},
  volume={107},
  number={4},
  pages={603--616},
  year={2020},
  publisher={Elsevier}
}

@article{jager2021using,
  title={Using neural networks for a universal framework for agent-based models},
  author={J{\"a}ger, Georg},
  journal={Mathematical and Computer Modelling of Dynamical Systems},
  volume={27},
  number={1},
  pages={162--178},
  year={2021},
  publisher={Taylor \& Francis}
}

@article{strnad2019deep,
  title={Deep reinforcement learning in World-Earth system models to discover sustainable management strategies},
  author={Strnad, Felix M and Barfuss, Wolfram and Donges, Jonathan F and Heitzig, Jobst},
  journal={Chaos: An Interdisciplinary Journal of Nonlinear Science},
  volume={29},
  number={12},
  year={2019},
  publisher={AIP Publishing}
}

@article{barfuss2019deterministic,
  title={Deterministic limit of temporal difference reinforcement learning for stochastic games},
  author={Barfuss, Wolfram and Donges, Jonathan F and Kurths, J{\"u}rgen},
  journal={Physical Review E},
  volume={99},
  number={4},
  pages={043305},
  year={2019},
  publisher={APS}
}

@article{zheng2022ai,
  title={The AI Economist: Taxation policy design via two-level deep multiagent reinforcement learning},
  author={Zheng, Stephan and Trott, Alexander and Srinivasa, Sunil and Parkes, David C and Socher, Richard},
  journal={Science advances},
  volume={8},
  number={18},
  pages={eabk2607},
  year={2022},
  publisher={American Association for the Advancement of Science}
}

@article{gershtenson2010creating,
  title={Creating better citizens? Effects of a model citizens' assembly on student political attitudes and behavior},
  author={Gershtenson, Joseph and Rainey Jr, Glenn W and Rainey, Jane G},
  journal={Journal of Political Science Education},
  volume={6},
  number={2},
  pages={95--116},
  year={2010},
  publisher={Taylor \& Francis}
}

@article{suiter2016deliberative,
  title={When do deliberative citizens change their opinions? Evidence from the Irish Citizens’ Assembly},
  author={Suiter, Jane and Farrell, David M and O’Malley, Eoin},
  journal={International Political Science Review},
  volume={37},
  number={2},
  pages={198--212},
  year={2016},
  publisher={Sage Publications Sage UK: London, England}
}

@article{kuntze2021citizen,
  title={Citizen assemblies can enhance political feasibility of ambitious climate policies},
  author={Kuntze, Lennart and Fesenfeld, Lukas Paul},
  journal={Available at SSRN 3918532},
  year={2021}
}

@article{king2023local,
  title={Local government and democratic innovations: reflections on the case of citizen assemblies on climate change},
  author={King, Martin and Wilson, Rob},
  journal={Public Money \& Management},
  volume={43},
  number={1},
  pages={73--76},
  year={2023},
  publisher={Taylor \& Francis}
}

@article{ejsing2023green,
  title={Green politics beyond the state: radicalizing the democratic potentials of climate citizens’ assemblies},
  author={Ejsing, Mads and Veng, Adam and Papazu, Irina},
  journal={Climatic Change},
  volume={176},
  number={6},
  pages={73},
  year={2023},
  publisher={Springer}
}

@article{golub2010naive,
  title={Naive learning in social networks and the wisdom of crowds},
  author={Golub, Benjamin and Jackson, Matthew O},
  journal={American Economic Journal: Microeconomics},
  volume={2},
  number={1},
  pages={112--149},
  year={2010},
  publisher={American Economic Association}
}

@article{banerjee1992simple,
  title={A simple model of herd behavior},
  author={Banerjee, Abhijit V},
  journal={The quarterly journal of economics},
  volume={107},
  number={3},
  pages={797--817},
  year={1992},
  publisher={MIT Press}
}

@book{russell1918proposed,
  author    = {Bertrand Russell},
  title     = {Proposed Roads to Freedom: Socialism, Anarchism and Syndicalism},
  year      = {1918},
  publisher = {George Allen \& Unwin},
  address   = {London},
  url       = {https://www.gutenberg.org/ebooks/690},  % optional: link to Project Gutenberg
  note      = {Reprinted by many publishers, including Routledge and Dover Publications.}
}

@book{plato2004republic,
  author    = {Plato},
  title     = {The Republic},
  year      = {2004},
  editor    = {C. D. C. Reeve},
  translator= {G. M. A. Grube},
  publisher = {Hackett Publishing Company},
  address   = {Indianapolis, IN},
  edition   = {2nd},
  url       = {https://www.gutenberg.org/ebooks/1497},  % optional: link to public domain version
  note      = {Originally written c. 380 BCE; various translations available.}
}

@article{brinkmann2023machine,
  title={Machine culture},
  author={Brinkmann, Levin and Baumann, Fabian and Bonnefon, Jean-Fran{\c{c}}ois and Derex, Maxime and M{\"u}ller, Thomas F and Nussberger, Anne-Marie and Czaplicka, Agnieszka and Acerbi, Alberto and Griffiths, Thomas L and Henrich, Joseph and others},
  journal={Nature Human Behaviour},
  volume={7},
  number={11},
  pages={1855--1868},
  year={2023},
  publisher={Nature Publishing Group UK London}
}

@book{mcneill1963rise,
  title={The Rise of the West: A History of the Human Community},
  author={McNeill, William H.},
  year={1963},
  publisher={University of Chicago Press},
  address={Chicago},
  isbn={978-0-226-56141-7}
}

@article{carballo2014cooperation,
  title={Cooperation and collective action in the cultural evolution of complex societies},
  author={Carballo, David M and Roscoe, Paul and Feinman, Gary M},
  journal={Journal of Archaeological Method and Theory},
  volume={21},
  pages={98--133},
  year={2014},
  publisher={Springer}
}

@book{conte1995cognitive,
  title={Cognitive and Social Action},
  author={Conte, Rosaria and Castelfranchi, Cristiano},
  year={1995},
  publisher={Psychology Press},
  address={London},
  isbn={9781857281866}
}

@article{wang2025limits,
  title={What Limits LLM-based Human Simulation: LLMs or Our Design?},
  author={Wang, Qian and Wu, Jiaying and Tang, Zhenheng and Luo, Bingqiao and Chen, Nuo and Chen, Wei and He, Bingsheng},
  journal={arXiv preprint arXiv:2501.08579},
  year={2025}
}

@article{hammond2025multi,
  title={Multi-Agent Risks from Advanced AI},
  author={Hammond, Lewis and Chan, Alan and Clifton, Jesse and Hoelscher-Obermaier, Jason and Khan, Akbir and McLean, Euan and Smith, Chandler and Barfuss, Wolfram and Foerster, Jakob and Gaven{\v{c}}iak, Tom{\'a}{\v{s}} and others},
  journal={arXiv preprint arXiv:2502.14143},
  year={2025}
}

@article{barfuss2022dynamical,
  title={Dynamical systems as a level of cognitive analysis of multi-agent learning: Algorithmic foundations of temporal-difference learning dynamics},
  author={Barfuss, Wolfram},
  journal={Neural Computing and Applications},
  volume={34},
  number={3},
  pages={1653--1671},
  year={2022},
  publisher={Springer}
}

@article{khamassi2024strong,
  title={Strong and weak alignment of large language models with human values},
  author={Khamassi, Mehdi and Nahon, Marceau and Chatila, Raja},
  journal={Scientific Reports},
  volume={14},
  number={1},
  pages={19399},
  year={2024},
  publisher={Nature Publishing Group UK London}
}
\end{document}